\documentclass{aa} 
\usepackage[varg]{txfonts}
\usepackage{graphicx}

\def\teff{$T\rm_{eff }$}

\newcommand{\fei}{Fe\,{\scriptsize I}}
\newcommand{\feii}{Fe\,{\scriptsize II}}

\def\feih{\hbox{[\rm{Fe\,{\scriptsize I}/H}]}}
\def\feiih{\hbox{[\rm{Fe\,{\scriptsize II}/H}]}}
\def\Rg{R\rm_G}

\begin{document}

\title{On the metallicity distribution of classical Cepheids in the Galactic
inner disk\thanks{Based on spectra collected with the spectrograph UVES available at the ESO Very Large Telescope (VLT), Cerro Paranal, (081.D-0928(A) PI: S.~Pedicelli -- 082.D-0901(A) PI: S.~Pedicelli)}}

\author{K. Genovali\inst{1}
\and B. Lemasle\inst{2}
\and G. Bono\inst{1,3}
\and M. Romaniello\inst{4}
\and F. Primas\inst{4}
\and M. Fabrizio\inst{5}
\and R. Buonanno\inst{1,5}
\and P. Fran\c cois\inst{6,7}
\and L. Inno\inst{1,4}
\and C.D. Laney\inst{8,9}
\and N. Matsunaga\inst{10}
\and S. Pedicelli\inst{1}
\and F. Th\'evenin\inst{11}}

\institute{Dipartimento di Fisica, Universit\`{a} di Roma Tor Vergata, via della Ricerca Scientifica 1, 00133 Rome, Italy \email{katia.genovali@roma2.infn.it}
\and Astronomical Institute `Anton Pannekoek', Science Park 904, P.O. Box 94249, 1090 GE Amsterdam, The Netherlands
\and INAF--Osservatorio Astronomico di Roma, via Frascati 33, Monte Porzio Catone, Rome, Italy
\and European Southern Observatory, Karl-Schwarzschild-Str. 2, D-85748 Garching bei Munchen, Germany
\and INAF-Osservatorio Astronomico di Collurania, via M. Maggini, I-64100 Teramo, Italy
\and GEPI - Observatoire de Paris, 64 Avenue de l'Observatoire, 75014 Paris, France
\and UPJV - Universit\'e de Picardie Jules Verne, 80000 Amiens, France
\and Department of Physics and Astronomy, N283 ESC, Brigham Young University, Provo, UT 84601, USA
\and South African Astronomical Observatory, P.O. Box 9, Observatory 7935, South Africa
\and Kiso Observatory, Institute of Astronomy, School of Science, The University of Tokyo 10762-30, Mitake, Kiso-machi, Kiso-gun,3 Nagano 97-0101, Japan
\and Laboratoire Lagrange, UMR7293, Universit\'e de Nice Sophia-Antipolis, CNRS, Observatoire de la C\^ote d'Azur, 06300 Nice, France
}
\date{Received <date> / Accepted <date>}

\abstract{
We present homogeneous and accurate iron abundances for almost four dozen (47) of Galactic Cepheids 
using high--spectral resolution (R$\sim$40,000) high signal--to--noise ratio (SNR $\ge$ 100) 
optical spectra collected with UVES at VLT. A significant fraction of the sample (32) is 
located in the inner disk ($\Rg \le$ 6.9 kpc) and for half of them we provide new iron 
abundances. Current findings indicate a steady increase in iron abundance when approaching 
the innermost regions of the thin disk. The metallicity is super--solar and ranges from 
0.2 dex for $\Rg\sim$6.5 kpc to 0.4 dex for $\Rg\sim$5.5 kpc. Moreover, we do not find 
evidence of correlation between iron abundance and distance from the Galactic plane. 
We collected similar data available in the literature and ended up with a sample of 
420 Cepheids. Current data suggest that the mean metallicity and the metallicity 
dispersion in the four quadrants of the Galactic disk attain similar values. 
The first--second quadrants show a more extended metal-poor tail, while the 
third--fourth quadrants show a more extended metal--rich tail, but 
the bulk of the sample is at solar iron abundance. 
Finally, we found a significant difference between the iron abundance of Cepheids 
located close to the edge of the inner disk  ([Fe/H]$\sim$0.4) and young stars 
located either along the Galactic Bar or in the Nuclear Bulge ([Fe/H]$\sim$0).   
Thus suggesting that the above regions have had different chemical enrichment 
histories. The same outcome applies to the metallicity gradient of the Galactic 
Bulge, since mounting empirical evidence indicates that the mean metallicity 
increases when moving from the outer to the inner Bulge regions.   }


\maketitle

\titlerunning{Cepheid metallicity distribution in the Galactic inner disk}
\authorrunning{Genovali et al.}

\section{Introduction}

Classical Cepheids after the discovery that they are robust distance indicators 
\citep{Leavitt1912} have had a widespread use in stellar astrophysics and in cosmology
\citep{Freedman2010}. They are very popular since they can be adopted as 
primary distance indicators 
\citep{Matsunaga2011a,Storm2011b,Freedman2012,Inno2013,Groenewegen2013}, 
as fundamental physics 
laboratory to constrain evolutionary and pulsation properties of intermediate-mass 
stars \citep{Bono2010,Pietrynski2010,Molinaro2011,Prada2012,Matthews2012,Neilson2012} 
and as stellar tracers of young stars in the thin disk 
\citep{Andrievsky2004,Lemasle2007,Lemasle2008,Romaniello2008,Pedicelli2009,Luck2011b} 
and in the Magellanic Clouds \citep{Storm2011b,Groenewegen2013,Haschke2012}.   

In particular, the iron and the $\alpha$-element abundance gradients across 
the Galactic disk are fundamental observables to constrain the chemical 
enrichment of disk stellar populations ~\citep{Luck2006,Lemasle2008,Luck2011a,Luck2011b}.
They also play a key role in constraining the physical assumptions adopted in
chemical evolution models ~\citep{Portinari2000,Chiappini2001,Cescutti2007}.
The most recent theoretical and empirical investigations brought forward 
three relevant open issues:

{\em Stellar tracers}-- Empirical evidence indicates that different
stellar tracers do provide different slopes. Metallicity gradients based
on Cepheids provide slopes of the order of $-0.05$~dex~kpc$^{-1}$
~\citep{Caputo2001,Andrievsky2002b,Luck2003,Kovtyukh2005,Luck2006,Yong2006,
Lemasle2007,Lemasle2008}. More recently,
~\cite{Pedicelli2009} using iron abundances for 265 classical Cepheids
--based either on high-resolution spectra or on photometric metallicity 
indices-- and Galactocentric distances ranging from $\Rg$ $\sim$5 to 
$\Rg$ $\sim$17 kpc found an iron gradient of $-0.051\pm0.004$~dex~kpc$^{-1}$. 
By using an even larger sample of more than 400 classical Cepheids, 
~\cite{Luck2011b} found a gradient of $-0.062\pm 0.002$~dex~kpc$^{-1}$.
A similar gradient was also found by ~\cite{Friel2002} by using a sample of
39 open clusters located between the solar circle and $\Rg$ $\sim14$ kpc,
namely $-0.06$~dex~kpc$^{-1}$. On the other hand,~\cite{Carraro2007}, by using
new metallicities for five old open clusters located in the outer disk
(12$\le$ $\Rg$ $\le$21 kpc) and the sample adopted by ~\cite{Friel2002},
found a shallower iron gradient: $-0.018$~dex~kpc$^{-1}$.
The slope of the metallicity gradient based on oxygen abundances of HII regions
--with Galactocentric distances ranging from 5 to 15 kpc-- is similar to the
slope based on Cepheids ($-0.04$~dex~kpc$^{-1}$,~\citealt{Deharveng2000}).
The difference between the different tracers might be due to the age difference
between the different tracers (young vs intermediate-age).

{\em Linear slope}-- By using a sample of 76 open clusters with distances
ranging from 6 to 15 kpc, it was suggested by ~\cite{Twarog1997} that a proper
fit to the metallicity distribution does require two zones. The inner disk for
Galactocentric distances ranging from 6 to 10 kpc and the outer disk for 
distances larger than 10 kpc. This hypothesis was supported by 
~\cite{Caputo2001}, \cite{Luck2003}, and \cite{Andrievsky2004}.
More recently, ~\cite{Pedicelli2009} found for the two zones a slope of
$-0.130\pm0.015$~dex~kpc$^{-1}$ for the inner disk ($\Rg < 8$ kpc) 
and a slope of $-0.042\pm0.004$~dex~kpc$^{-1}$ for the outer disk.

{\em Local inhomogeneities}-- There is evidence of
abundance inhomogeneities across the Galactic quadrants. 
~\cite{Pedicelli2009} found that the iron abundance of the two Cepheid overdensities located in the
first and in the third quadrant covers the same metallicity range of the global gradient (see also
\citealt{Luck2006,Lemasle2007,Luck2011a}).

{\em Azimuthal dependence}-- In a recent investigation concerning the 
chemical enrichment of the thin disk based on 398 Cepheids, \cite{Luck2011b}
did not find evidence of azimuthal dependence. \cite{Davies2009b}    
suggested that the increase in metallicity in region located between the solar 
circle and the inner disk might be due to an azimuthal dependence. 

The above open issues rely on sizable samples of Galactic classical Cepheids, 
but they are partially hampered by the fact that abundances based on high-resolution 
spectra are far from being homogeneous. To overcome these thorny problems, we 
collected a large sample of high spectral--resolution (R$\sim$38,000) spectra 
with UVES at VLT for 77 classical Cepheids mainly located in the third and 
fourth quadrant of the Galactic disk.   


\section{Spectroscopic data and iron abundances}\label{sec:metallicity}

We acquired high-resolution (R$\sim$38,000) spectra
with the UVES spectrograph 
available at the Nasmyth B focus of UT2/VLT 8m telescope (Fig. 1). 
The spectra were collected at random pulsational phases between 2008 October and 2009 April using the DIC2 (437+760) configuration  with blue and red arms operating in parallel. The two arms 
cover the wavelength intervals $\sim$3750--5000 \AA \ and 
$\sim$5650--7600/7660--9460 \AA \ (two chips in the red arm).

The complete spectroscopic sample includes 84 spectra for a total of 77 Cepheids.  
They were collected with exposure times ranging from  80 to 2000 sec, while 
the seeing ranges from 0.6 to 2 arcsec with a mean value of 1.2 arcsec. 
The quality of the spectra is very good, and indeed the signal--to--noise--ratio 
(SNR) is typically better than $\sim$100 for all the echelle orders. 
The spectra were reduced using the ESO UVES dedicated pipeline REFLEX v2.1
\citep{Ballester2011}. In this investigation we focus our attention on 
a sub-sample --47 out of the 77 targets-- a more detailed description 
of the data reduction will be discussed in a forthcoming paper.

We implemented a semi-automatic procedure to determine the continuum and to 
fit with a Gaussian the single line profile. The blended lines were fit with 
the sum of two Gaussians. We adopted the iron linelist provided by 
\cite{Romaniello2008} and typically we measured the equivalent width (EW) of $\sim$250 \fei 
\ and $\sim$40 \feii \ lines. Some of these lines are heavily blended in our 
spectra and to overcome thorny problems in the estimate of the mean iron 
abundance, they were neglected in the estimate of the abundance. For the 
spectra lacking of useful \feii \ (i.e. with EWs smaller than $\sim$150 \AA) 
we also measured the \feii \ lines listed in \cite{Pedicelli2010}. 
The final number of adopted lines is $\sim$230 for \fei \ and $\sim$55 
for \feii, respectively (see Table \ref{tab:tab_linelist}). 

\begin{table}
\caption{\fei \ and \feii \ linelist adopted 
to estimate iron abundances.}
\label{tab:tab_linelist}
\centering
\begin{tabular}{c l c c}
\hline
\hline
$\lambda$[\AA]  & Ion  & EP  & log\textit{gf}  \\
\hline
4893.82 & \feii & 2.83 & -4.45 \\ 
4917.23 & \fei & 4.19 & -1.29 \\ 
4923.93 & \feii & 2.88 & -1.35 \\ 
4924.77 & \fei & 2.28 & -2.24 \\ 
4932.08 & \fei & 4.65 & -1.49 \\ 
  ... & ... & ... & ... \\
\hline
\end{tabular}
\tablefoot{Iron line list, from left to right the columns display wavelength, ion identification, excitation potential (EP) and log\textit{gf} values. The table is available in its entirety via the link to the machine-readable version. A portion is shown here for guidance regarding its form and content.}
\end{table}

We updated the atomic data (oscillator strength and excitation potential) 
of the lines according to the values available in VALD (Vienna Atomic Lines 
Database\footnote{www.astro.uu.se/$\sim$vald/php/vald.php}, \citealt{Kupka2000}).

Iron abundances were estimated using MARCS atmospheric models \citep{Gustafsson2008}. Once we estimated the atmospheric parameters 
and provided a preliminary estimate of the metallicity, we performed an 
interpolation over a grid of LTE, plane-parallel models by using the 
dedicated code for abundance determination, \textit{fitline}, developed 
by one of the co-authors (P. Fran\c cois). 
For each spectrum we computed the curves of growth, for both neutral and 
ionized iron together with the plots showing $\feih$ vs excitation potential 
and $\feih$ vs EWs. The process is iterated until a good match between 
predicted and observed equivalent widths and metallicity is approached. 

The  effective temperature --\teff-- for individual spectra was independently 
estimated by using the line depth ratios (LDRs) method \citep{Kovtyukh2000}. 
Typically, we measured two dozen LDRs per spectrum. The surface gravity 
--$\log g$-- is derived by imposing the ionization balance between 
\fei \ and \feii. The individual micro-turbulent velocities --$v_{t}$-- 
were estimated by minimizing the $\feih$/EW slope, together with a 
visual inspection of the curves of growth. 

The maximum EW value included in the metallicity determination varies 
according to the metallicity itself and on the atmospheric parameters 
of the star. For a significant fraction of our spectra we were able to 
use only relatively weak lines (EW $\textless$ 120 m\AA) located along 
the linear part of the curve of growth. In a few cases the spectra were 
lacking of a significant number of weak lines (less than two \feii \ lines), 
therefore, we included in the analysis also lines with EW $\le$ 180 m\AA. 
The latters cause a mild increase in the uncertainties affecting the 
atmospheric parameters. 

The mean iron abundances given in Table \ref{tab:tab_distances} are the 
weighted mean of \feih \ and \feiih \ lines. The errors listed in column six 
are the standard deviations of the weighted mean, i.e. 
$\sigma_{Fe}$=$\sqrt{\sigma_{\fei}^2+\sigma_{\feii}^2}$, 
where $\sigma_{\fei}$ and $\sigma_{\feii}$ are the standard deviations 
of \feih \ and \feiih \ lines.
The iron abundances were estimated assuming the solar iron abundance 
provided by \cite{Grevesse1996}, i.e. A(Fe)${_\odot}$ = 7.50.

\section{Photometric data and distance estimates}\label{sec:distances}

To provide a homogeneous sample of Galactocentric distances ($\Rg$), we adopted 
near infrared (NIR) photometric data together with the reddening-free 
Period-Wesenheit relations in $J,H,K_s$ bands provided by \cite{Inno2013}. 
We estimated individual distances for a significant fraction of Galactic 
Cepheids (Genovali et al. 2013, in preparation). 
The NIR photometric catalogs we adopted are the following: 

{\em SAAO sample}-- The bulk of this data set comes from \citealt{Laney1992}.
The individual measurements provide an accurate coverage of the light curve  
and the typical accuracy of mean $J,H,K_s$ magnitudes is at the millimag level.    
This data set was complemented with new NIR measurements (Laney 2013, private 
communication) and includes objects with either complete or partial coverage 
of the light curve. For the latter group the number of phase points along the 
light curve ranges from four to 14. Their mean magnitudes were estimated 
using a spline. The SAAO NIR magnitudes were transformed into the 2MASS 
photometric system using the transformations provided by \cite{Koen2007}.     

{\em Monson \& Pierce sample}-- In a recent investigation \cite{Monson2011} 
published accurate NIR magnitudes for 131 northern hemisphere Cepheids. Their 
NIR mean magnitudes were transformed into the 2MASS photometric system using 
the transformations provided by the same authors. The measurements properly 
cover the entire pulsation cycle and their typical accuracy is better than 
0.01 mag.

{\em 2MASS sample}-- The above samples were complemented with 2MASS 
single-epoch observations \citep{2MASS}. The mean NIR magnitude of fundamental 
(FU) Cepheids were estimated by using single-epoch photometry and the light-curve 
template provided by \cite{Soszynski2005}. The pulsation properties required to 
apply the template (epoch of maximum, optical amplitudes, periods) come 
from General Catalog of Variable Stars
\citep[GCVS\footnote{\url{http://www.sai.msu.su/gcvs/gcvs/index.htm}};][]{Samus2009}. 
The error on the mean NIR magnitudes was estimated as 
$\sigma_{J,H,K_s}^2 = \sigma_{phot}^2+ \sigma_{temp}^2$, 
where $\sigma_{phot}$ is the intrinsic photometric error --typically 
of the order of 0.03 mag for the Cepheids in the current sample--  
and $\sigma_{temp}$=0.03 mag is the uncertainty associated with the 
intrinsic scatter of the template\footnote{Current sample includes 
only two first overtones --EV Sct, QZ Nor-- and their mean magnitudes 
were provided by \cite{Laney1992}.}.

For the objects in common in the three different samples, we adopted 
among them the most accurate. The individual distances were estimated 
as weighted mean of the three different distances obtained by adopting the NIR (\textit{H,J-H}; \textit{K,J-K}; \textit{K,H-K}) Period-Wesenheit (PW) relations  
provided by Inno et al. (2013). The individual distance moduli are 
listed in column 7 of Table \ref{tab:tab_distances} with their 
uncertainties.
The Galactocentric distances listed in column 8 of Table 
\ref{tab:tab_distances} were estimated assuming a solar Galactocentric 
distance of 7.94$\pm$0.37$\pm$0.26 kpc \citep[][and references therein]{Groenewegen2008, 
Matsunaga2013}. 
The final error on $\Rg$ accounts for errors affecting both the solar 
Galactocentric distance and the heliocentric distances. 

We compared current individual distances based on NIR PW relation with individual distances
estimated using two different flavors of the IRSB method and we found that the mean difference
over the entire sample ranges from 8$\pm$2\%  (Groenewegen 2013, $\sim$130 stars in common) 
to  3$\pm$2\% (Storm et al. 2011a, $\sim$80 stars in common).

\section{The metallicity gradient}

To further constrain the nature of the metallicity gradient of the Galactic disk, 
in Fig. 2 we plotted both new (16, red circles) and updated (31, blue circles) Cepheid iron 
abundances as a function of the Galactocentric distances ($\Rg$).   
Together with the current sample, we also included iron abundance for Galactic 
Cepheids estimated by our group using the same approach and similar data 
(\citealt{Lemasle2007,Lemasle2008}; 57, [LEM], green circles; 
\citealt{Romaniello2008}; 16, [ROM], yellow circles). Moreover, we also 
included Cepheid iron abundances available in the literature 
(\citealt{Sziladi2007,Luck2011a,Luck2011b}; 300, [LUCII, LUCIII], black circles).   
We ended up with a sample of 420 Cepheids. The different sets of abundances 
were re-scaled to the same metallicity 
scale\footnote{The difference in iron abundance among the different samples are: 
$\Delta [Fe/H]$(ROM-us)$=-0.06\pm 0.10$,
$\Delta [Fe/H]$(LEM-us)$=-0.11\pm0.09$, 
$\Delta [Fe/H]$(LUCII-us)$=-0.07\pm0.13$ and   
$\Delta [Fe/H]$(LUCIII-us)$=0.04\pm0.08$. The difference with the seven Cepheids 
by \cite{Sziladi2007} was not estimated, since we only have one object in common.}. 

The current sample has two indisputable advantages: 
a) iron abundances are based on high--resolution spectra and they were rescaled 
to the same solar abundance adopted in \S \ref{sec:metallicity};
b) distance estimates are homogeneous and based on PW relations that are independent 
of reddening uncertainties and minimally affected by metallicity dependence 
\citep{Inno2013}.  

Data plotted in Fig. \ref{fig:grad} show that the metallicity gradient shows a large 
intrinsic dispersion around the solar circle with the possible occurrence either of 
a change in the slope or of a shoulder for $7\le\Rg\le$10 kpc \citep{Andrievsky2004}. 
This finding supports 
previous results based on open clusters and on Cepheids obtained by 
Twarog et al. (1997) and by Caputo et al. (2001). However, a quantitative 
analysis of this issue does require a large sample of homogeneous and accurate iron 
abundance across the solar circle.   
The metallicity gradient also shows an increase in the intrinsic dispersion 
in the outer disk  ($\Rg\ge$ 13 kpc, $\sigma$(Fe/H)$\sim$0.15). 
This finding supports previous results by \cite{Pedicelli2009} and by \cite{Luck2011a}. There is also a mild evidence of a flattening, but 
the number of outer disk Cepheids is too limited to constrain the possible 
occurrence of a change in the slope \citep{Carraro2007}.  
Moreover, Cepheids located in the inner disk ($\Rg\le$ 6.9 kpc) show a well 
defined steepening in the gradient \citep{Andrievsky2002a,Pedicelli2010} and a steady trend concerning the intrinsic 
dispersion ($\sigma$(Fe/H)$\sim$0.1). 
It is worth mentioning that the current sample increases by 20\% the number of inner 
disk Cepheids with spectroscopic abundances and together with the updated iron 
abundances we are providing homogeneous abundances for the 55\% of inner disk 
Cepheids.  

The zoom on the region located between the Galactic center and the inner disk 
(bottom panel of Fig.~2) shows that Cepheids located in the inner disk attain 
super-solar iron abundances. The trend is quite clear down to $\Rg\sim$ 5 kpc, 
but becomes less clear towards the edge of the inner disk. Indeed, only two 
Cepheids in our sample have smaller Galactocentric distances. Current findings 
support previous results by \cite{Andrievsky2002a}, Pedicelli et al. (2009)
and by \cite{Luck2011a}.       

Although current knowledge of the innermost components of the Galactic 
spheroid --Nuclear Bulge, Galactic Bulge, inner disk-- is quite solid 
~\citep{Brand1993,Honma1997,Reid2009},  we still lack quantitative constrains 
on their kinematic structure and on their chemical enrichment history. 
In particular, current predictions suggest that a presence
of a bar-like structure is crucial to support the high star formation rate 
of the Nuclear Bulge \citep{Yusef2009,Davies2009a,Matsunaga2011b}. 
As a matter of fact, it is the bar-like structure to drag the gas and 
the molecular clouds from the inner disk into the Nuclear Bulge 
~\citep{Athanassoula1992,Kim2011}. 

To further constrain this crucial issue we plotted in Fig.~2 the iron 
abundances for young stars located in the Nuclear Bulge. The red cross 
shows the mean iron abundance of two Luminous Blue Variables (LBVs) in 
the Quintuplet cluster provided by ~\cite{Najarro2009}. The green 
triangle shows the mean iron abundance of a dozen nitrogen Wolf-Rayet 
stars (WNLh) in the Arches cluster provided by \cite{Martins2008}, 
while the green square marks the iron abundance of two field red 
supergiants (RSGs) of the Nuclear Bulge provided by 
~\cite{Davies2009a}. The main outcome of these investigations is that 
the typical iron abundance of both cluster and field young stellar 
tracers is solar.  This empirical evidence is further supported by 
recent spectroscopic measurements provided by \cite{Davies2009b} 
for two Scutum RGS clusters (14 and 12 stars) located at the near 
end of the Galactic Bar (see magenta diamonds in Fig.~2). They found 
that both of them have sub-solar iron abundance (--0.2- --0.3 dex).       

The above findings indicate that the iron abundance of stellar 
tracers located along the Galactic Bar and in the Nuclear Bulge are
more metal-poor than classical Cepheids in the inner disk. This 
appears as a solid evidence, since the difference in age among the 
different tracers is minimal, i.e. from a few Myr to a few tens of 
Myrs. Thus suggesting that the mechanism(s) feeding of gas the 
Bar and the Nuclear Bulge might be more complex than currently 
assumed.  

It has also been suggested that the Nuclear Bulge could have been 
formed by the Galactic Bulge \citep{Morris1996,Davies2009a,Davies2009b}. 
To further constrain this fundamental issue, we 
plotted in Fig.~2 the mean metallicity of four Galactic Bulge fields 
recently provided by \cite{Zoccali2008} (l=0$^{\circ}$, b=$-4^{\circ},
-6^{\circ},-12^{\circ}$) and by \cite{Uttenthaler2012} 
(l=0$^{\circ}$, b=$-10^{\circ}$). By using high--resolution spectra for 
a large sample of field Red Giant and Red Clump 
stars (521), \cite{Zoccali2003,Zoccali2008} and \citet{Hill2011} found that the metallicity 
distribution of Galactic Bulge stars shows a broad main peak [Fe/H]$\approx$-0.15
and a large spread in metallicity with iron abundances ranging (1$\sigma$ variation)
from metal-rich ([Fe/H]$\sim$-0.55) to super solar ([Fe/H]$\sim$+0.25, 
see the vertical solid gray line in Fig.~2). 
The same outcome applies to the sample of 383 RC and RG stars measured by 
\cite{Uttenthaler2012}, with iron abundances ranging (1$\sigma$ variation) 
from [Fe/H]$\sim$-0.85 to $\sim$0.17 (see the vertical dashed gray line in Fig.~2).   
Unfortunately, we lack detailed information concerning the Galactocentric 
distances of the above Bulge fields, however, plain geometrical arguments 
concerning their Galactic coordinate indicate that, on average, the metallicity 
gradient is steadily increasing when moving from the outer to the inner edge 
of the Galactic Bulge \citep{Ness2013}.
This appears a solid finding, since they adopted old and intermediate-age 
stellar tracers, i.e. the typical stellar populations of the Galactic Bulge.  

Once again, the metallicity distributions in the inner disk, in the Galactic 
Bulge, in the Nuclear Bulge and in the Galactic Bar display different trends.   
Plain leading arguments based on the extrapolation of the inner disk 
metallicity gradient would imply super-solar iron abundances approaching 
the Galactic Center~\citep{Davies2009a}. This discrepancy was already noted 
in the literature and ~\cite{Cunha2007} suggested that the slope of the 
metallicity gradient should become shallower for Galactocentric distances 
smaller than 5~kpc. On the other hand, current findings indicate that young 
stellar tracers attain super-solar iron abundances approaching the inner 
edge of the thin disk.

To further constrain the metallicity distribution across the thin disk 
we also investigated the distribution of Galactic Cepheids onto the 
Galactic plane. The right panel of Fig.~3 shows the same objects 
plotted in Fig.~2, but the symbol size scales with the iron 
abundance (see labelled values). 
Current Cepheid sample (47) allowed us to improve the spatial 
sampling in the third and in the fourth quadrant.  
We now have roughly one hundred Cepheids per quadrant and their mean 
metallicities and metallicity dispersions attain similar values.   
There is evidence of an increase in the mean metallicity 
when moving from the first--second to the third-fourth quadrants, 
but the difference is on average smaller than 0.2 dex. 
To provide more quantitative constraints on the recent chemical 
enrichment history across the thin disk, we estimated the metallicity 
distribution of the four quadrants. Data plotted in the right 
panel of Fig.~3 show both the histograms (black line) and the 
smoothed distributions (red dotted line) of the iron abundances. 
The latter was estimated by running a Gaussian kernel with a 
mean $\sigma$ of 0.1 dex. 
The adopted $\sigma$ accounts for errors on individual
abundances and for uncertainties among different samples.   
The mean and the $\sigma$ of the Gaussian fit of the smoothed 
distributions are labelled. Data plotted in the right panel 
show, once again, that the difference is on average smaller than 
0.2 dex, moreover, the $\sigma$ attain very similar values.    

The above findings indicate that the chemical enrichment across the 
Galactic disk has been quite homogeneous during the last $\approx$100 Myr.
There is evidence of a more extended metal--poor tail in the first--second
quadrant and a more metal--rich tail in the third--fourth quadrant, 
but the bulk of the sample is at solar iron abundance.  
There is a mild evidence that the metallicity distribution is 
probably approaching a plateau toward the shortest Galactocentric 
distances  ($\Rg\le$ 5.5 kpc, [Fe/H]$\approx$0.4). However, this 
preliminary evidence is hampered by the limited number of Cepheids 
located close to the edge of the inner disk and by the lack of firm 
empirical estimates of the transition between the disk and the Bulge.   

In their recent investigation concerning the metallicity 
distribution of the Galactic Bar and of the Nuclear Bulge,  
\cite{Davies2009b} suggested that the increase in the mean metal 
abundance of inner disk Cepheids might be due to azimuthal variations. 
To constrain a possible change in the azimuthal direction, Fig.~4 shows 
the distance from the Galactic plane of entire Cepheid sample versus 
Galactocentric radius. 
Note that the Cepheid distribution shows a shift compared with the 
Galactic plane similar to the shift originally found by \cite{Kraft1964} 
and by \cite{Fernie1995}, i.e. $\Delta Z=-43\pm13$ pc (420 stars). 
It is worth mentioning that the distribution of azimuthal distances 
remain roughly constant up to ten kpc. At larger Galactocentric distances 
the dispersion increases and shows a well defined departure for $\Rg$ 
larger than 13 kpc.  
To constrain a possible metallicity dependence, we split the sample in 
three sub-samples according to the metallicity and we found that the 
difference in azimuthal distance among them is smaller than 1$\sigma$.

\section{Conclusions and final remarks}

We performed accurate measurements of iron abundances for 47 Galactic 
Cepheids using high-resolution, high SNR spectra collected with UVES 
at VLT. The new sample includes 32 out of the 58 classical Cepheids located in the inner disk ($\Rg$ $\le$ 6.9 kpc) for which are available accurate iron abundances. 

We found that the metallicity distribution of the inner disk 
Cepheids is super-solar, and ranges from  $\sim0.2$ dex 
($ \Rg \approx 6.5$ kpc) to $\sim0.4$ dex ($\Rg\approx 5.5~kpc$). 
Current finding supports previous results \citep{Pedicelli2009,Luck2011b} 
concerning a steady increase in metallicity when approaching the innermost 
disk regions. The new sample allowed us to double the number of Cepheids in 
the third and in the fourth quadrants when compared with \cite{Pedicelli2009}. 
We found that the difference in the mean metallicity is smaller than 0.2 dex 
and the metallicity dispersion attain quite similar values in the four quadrants. 
The metallicity appears to approach a plateau value close to the edge of the 
inner disk, but this evidence is hampered by the limited number of known Cepheids 
with $\Rg\lesssim 5~kpc$. In this context, it is worth mentioning that the 
current sample includes at least the 80\% of the known classical Cepheids 
located in the inner disk (62$+$7 candidates, Genovali et al. 2013, in preparation).  
 
Moreover, we do not find evidence of an azimuthal variation 
in the Cepheid metallicity distribution. This finding supports previous 
results by \cite{Luck2011b}.
The Cepheids close to the edge of the inner disk 
($\Rg$ $\approx$ 4 kpc) are significantly more metal--rich than young 
stellar tracers located along the Galactic Bar and in the Nuclear Bulge. 
The difference is of the order of 0.5 dex, since Cepheids attain iron abundances 
of 0.4--0.5 dex while LBVs and red supergiants have either solar or sub--solar  
metallicities. This evidence indicates that young stars at the inner edge and 
at the Galactic Bar/Nuclear Bulge formed from material far from being 
homogeneous concerning the chemical enrichment.  

Theoretical ~\citep{Athanassoula1992,Friedli1995} and empirical  
\citep{Zaritsky1994,Allard2006,Zanmar2008} investigations indicate that
the abundance gradient in barred galaxies is shallower than in 
unbarred galaxies. The typical explanation for this trend is that 
the bar is dragging gas from the inner disk into the Nuclear Bulge 
~\citep{Kim2011}. The pileup of the new fresh material triggers
an ongoing star formation activity till the dynamical stability 
of the bar.
However, there is mounting evidence that in our galaxy the formation 
and evolution of the above components might have been more complex
\citep{Ness2012}. 
This evidence is further supported by the fact that we are using 
either massive or intermediate--mass stars (Supergiants, LBVs, 
Cepheids) and their evolutionary lifetime is typically shorter 
than 100 Myr.   
This indicates that the bar might not be the main culprit
in shaping the metallicity gradient between the inner disk 
and the Nuclear Bulge/Galactic Bar. Indeed, recent numerical 
simulations suggest that the timescale within which the radial 
motion of the gas smooths the actual abundance gradient is of 
the order of a few hundred Myrs \citep{Merlin2007}. In this 
timescale part of the azimuthal variations observed across 
the Galactic disk might be caused
by changes in the abundance patterns between the spiral arms and the
inter-arm regions ~\citep{Kim2011} and by the clumpiness of the star
formation episodes. However, the kinematics of the above stellar tracers
is quite limited, since they evolve in situ. This working 
hypothesis is supported by the minimal dependence of the metallicity 
distribution from the azimuthal distance.  

The hypothesis suggested by ~\cite{Davies2009a} that the wind of
metal-intermediate bulge stars might mix with metal-rich gas present
along the Bar and the Nuclear Bulge to produce a chemical mixture
close to solar appears also very promising. However, recent 
spectroscopic investigations support the evidence of a metallicity 
distribution that is steadily increasing when moving from the outer 
to the inner edge of the Galactic Bulge \citep{Zoccali2008,Uttenthaler2012}. 
Moreover, we still lack firm empirical evidence on how the winds 
of bulge stars might fall into the Nuclear Bulge. 

Finally, it is worth mentioning that the infall of metal-poor gas in the
Nuclear Bulge appears an even more plausible channel to explain current
abundance patterns ~\citep{Wakker1999,Lubowich2000}. This is the so-called
biased infall scenario ~\citep{Chiappini1999} in which the
infall of gas takes place more rapidly in the innermost than in the
outermost regions (inside-out disk formation).

The abundance pattern of $\alpha$-elements is even more puzzling,
since accurate measurements indicate solar abundances, and therefore
consistent with typical thin disk stars ~\citep{Davies2009b}.
However, different tracers (B-type stars, red supergiants, classical
Cepheids, HII regions) do provide slightly different mean values,
suggesting a broad distribution (\citealt{Davies2009a}, and references therein).

A more quantitative understanding of the above phenomena is important 
not only to trace the chemical enrichment of the innermost Galaxy 
regions, but also for its impact on the formation and the evolution 
of classical bulges and pseudo-bulges 
~\citep{Kormendy2004,Kormendy2009,Matsunaga2011b,Freeman2013}.

This is the sixth paper in a series focussed on the spectroscopic abundances 
of Galactic Cepheids and together with similar investigations available in the 
literature we are approaching a complete spectroscopic census of known Galactic 
Cepheids. However, the current sample is severely affected by an observational bias both in the inner and in the outer disk direction. The new ground-based 
optical (Large Synoptic Survey Telescope, \citealt{Chang2013}) and NIR (IRSF/SIRIUS, \citealt{Nishiyama2005}) photometric surveys together with Gaia will fill this gap.

\begin{acknowledgements}
It is a pleasure to thank the referee, R. Earle Luck, for his positive opinion concerning the content and the cut of our paper.
This work was partially supported by PRIN–INAF 2011 ``Tracing the formation 
and evolution of the Galactic halo with VST'' (PI: M. Marconi) and by 
PRIN–MIUR (2010LY5N2T) ``Chemical and dynamical evolution of the Milky Way 
and Local Group galaxies'' (PI: F. Matteucci).
Three of us (G.B., K.G., M.F.) thank ESO for support as science visitors.
This publication makes use of data products from the Two Micron All Sky Survey, 
which is a joint project of the University of Massachusetts and the Infrared 
Processing and Analysis Center/California Institute of Technology, funded by 
the National Aeronautics and Space Administration and the National Science 
Foundation.
During the writing of this manuscript we learnt that A.N. Cox passed away. We dedicate this manuscript to his memory. His many contributions to the field of stellar pulsation, his steady support to young researchers and his captivating
enthusiasm will be greatly missed.
\end{acknowledgements}


\clearpage

\begin{figure*}
\centering
\includegraphics[width=17cm]{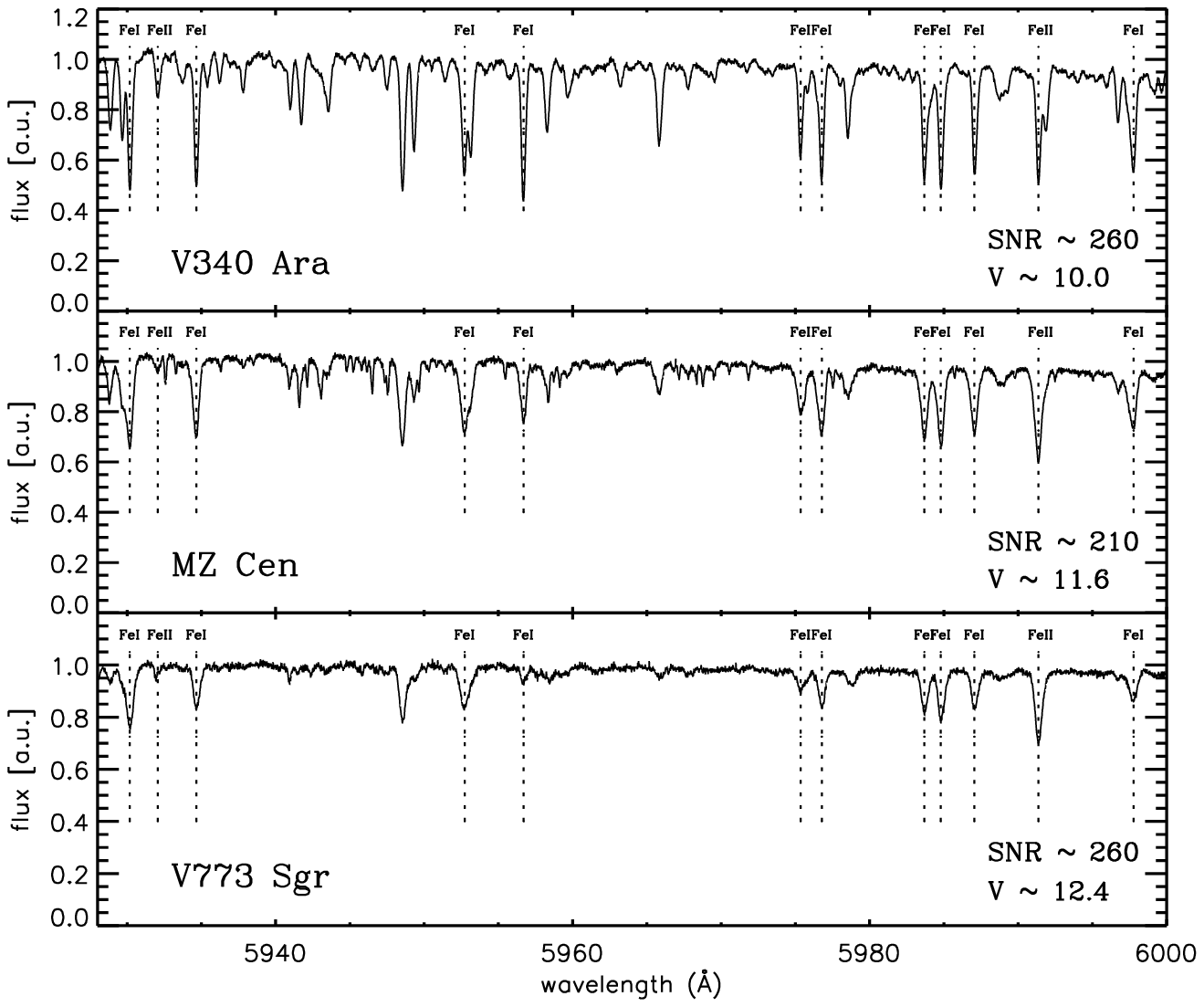}
\caption{Selected high-resolution (R$\sim$38,000) UVES spectra  of Galactic 
Cepheids in our sample. From top to bottom are plotted the spectra of 
V340 Ara ([Fe/H]=0.53$\pm$0.09), MZ Cen ([Fe/H]=0.27$\pm$0.10) and 
V773 Sgr ([Fe/H]=0.11$\pm$0.06). The apparent visual magnitude and the SNR 
in the spectral range $\lambda$ $\sim$5650--7500 \AA \ (red arm) are also labelled. 
The vertical dashed lines display selected \fei 
($\lambda\lambda$5930.17, 5934.66, 5952.73, 5956.7, 5975.35, 5976.78, 
5983.69, 5984.79, 5987.05, 5997.78 \AA) and \feii \ 
($\lambda\lambda$5932.06 and 5991.37 \AA) lines included in our 
abundance analysis \citep[see Table~1 and][]{Romaniello2008}. \label{fig:spettri}}
\end{figure*}

\begin{figure*}
\centering
\includegraphics[width=17cm]{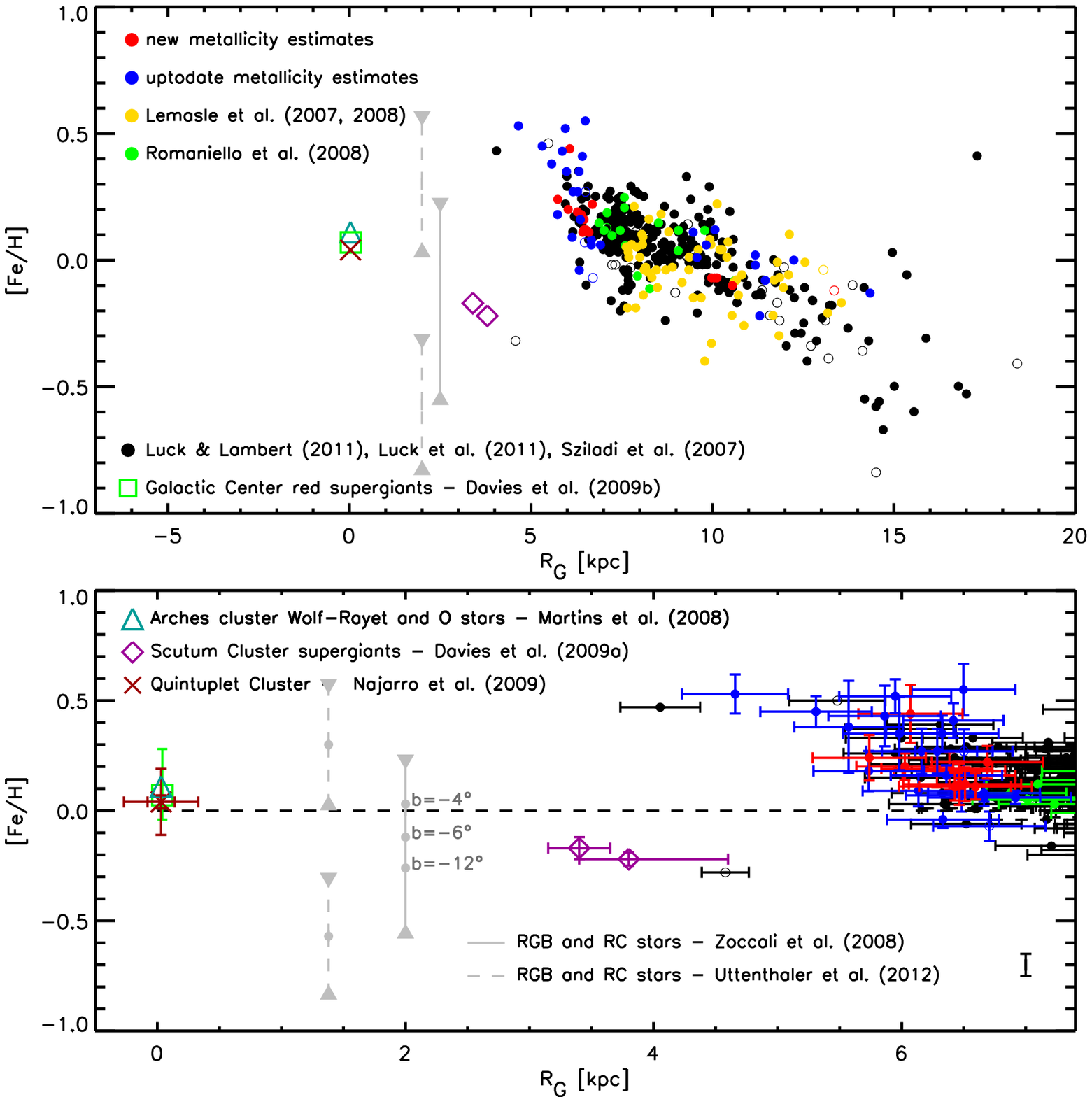}
\caption{Top-- Iron abundance of Galactic classical Cepheids versus 
Galactocentric distance. Spectroscopic measurements based on different 
data sets are plotted with different colors. New iron abundances based 
on current UVES spectra (16, red), update iron abundances based on 
UVES spectra (31, blue);  iron abundances 
provided by our group: \cite{Lemasle2008}, \cite{Lemasle2007} 
(57, yellow), \cite{Romaniello2008} (16, green); 
iron abundances available in the literature: \cite{Luck2011a}, 
\cite{Luck2011b}, and \cite{Sziladi2007} (300, black).
Cepheids that according to the General Catalog of Variable Stars 
\citep{Samus2009} are candidate classical Cepheids were plotted 
with open circles.  
The vertical gray bars located at arbitrary Galactocentric distances  
--$\Rg$ $\sim$ 2 kpc-- show the mean ($\pm$1$\sigma$) iron abundance 
for Galactic Bulge stars: the solid line is located at the mean 
metallicity of the sample provided by \cite{Zoccali2008}; 
the dashed lines are located at the mean values ($\pm$1$\sigma$) 
found by \cite{Uttenthaler2012}. The range in metallicity is 1$\sigma$. 
The green square marks the iron abundance of the two red supergiants 
in the Galactic center measured by \citep{Davies2009a}, while the magenta 
diamonds the 26 red supergiants in the Scutum Cluster measured by 
\citep{Davies2009b}, the red cross the two luminous blue variables 
(LBVs) in the Quintuplet cluster \citep{Najarro2009} and the 
light-blue triangle three Wolf--Rayet and two O-type stars in the 
Arches cluster \citep{Martins2008}.
Bottom--  Same as the the top, but for Galactocentric distances smaller than  
the solar circle ($\Rg$ $\leq$ 7.5 kpc). The bars on individual Cepheids display 
the uncertainty on both iron abundance and distance. The vertical black 
bar on the bottom right corner shows the mean uncertainty on 
\cite{Luck2011b} and \cite{Romaniello2008} iron abundances. The gray vertical 
lines are the same of the top panel; the filled circles on the solid line display 
the mean iron abundance of the bulge fields located at 
b=--4$^{\circ}$, ($\langle$[Fe/H]$\rangle$ = +0.03$\pm$ 0.38), 
b=--6$^{\circ}$, ($\langle$[Fe/H]$\rangle$ = --0.12$\pm$ 0.35), 
and at b=--12$^{\circ}$, ($\langle$[Fe/H]$\rangle$ = --0.26$\pm$ 0.40)
observed by \cite{Zoccali2008}. The filled circles on the dashed lines display 
the mean metallicities provided by \cite{Uttenthaler2012} for stars located at 
b=--10$^{\circ}$, ($\langle$[Fe/H]$\rangle$ = --0.57$\pm$ 0.27 and 
$\langle$[Fe/H]$\rangle$ = +0.30$\pm$ 0.28).
}\label{fig:grad}
\end{figure*}

\begin{figure*}
\centering
\includegraphics[width=9cm]{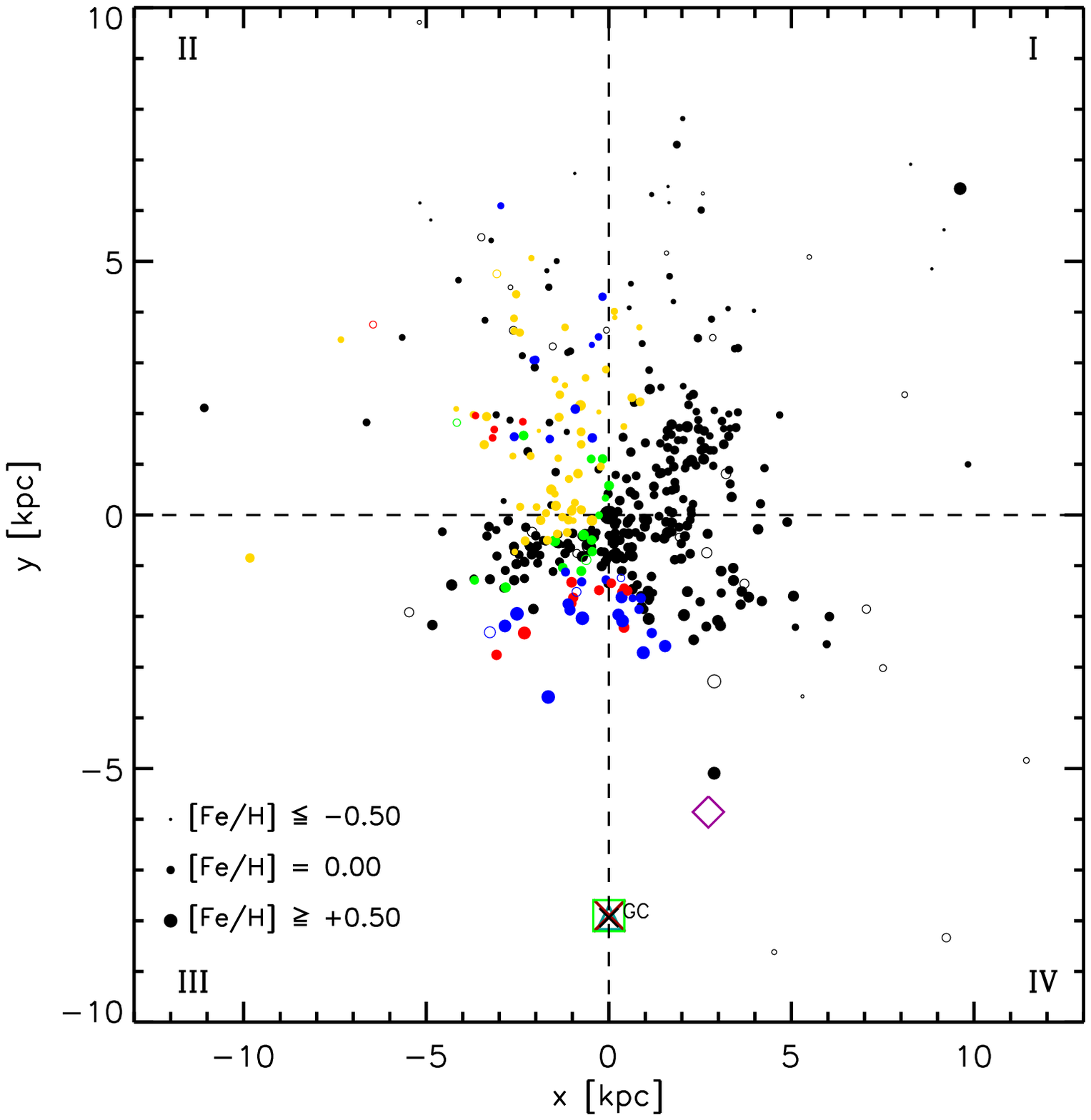}
\includegraphics[width=9cm]{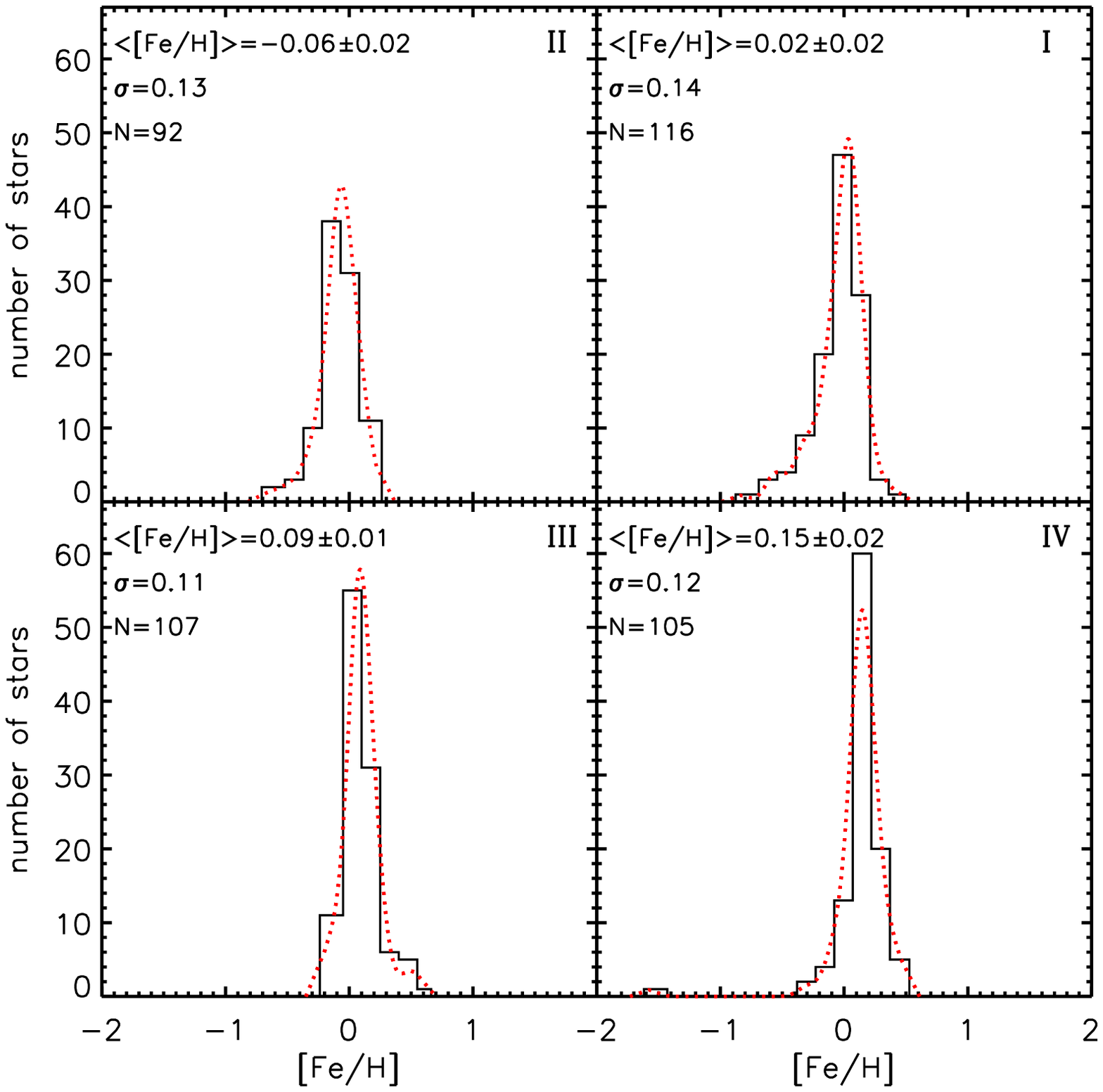}
\caption{
Left -- Distribution onto the Galactic plane of Cepheids with accurate iron 
abundances. Symbols and colors are the same as in Fig.~2. The size of the 
circles scales with their iron abundances (see labelled values). The bulge fields 
were not plotted, since we lack individual distances. The dashed lines display 
the position of the Sun, while the cross the position of the Galactic center. 
Right -- Metallicity distribution of the Cepheids in the four quadrants. The 
solid black line shows the histogram (bin size=0.15 dex). The dotted red line 
shows the smoothed metallicity distribution estimated running a Gaussian kernel 
with fixed $\sigma$=0.1 dex. The adopted dispersion accounts for errors on 
abundances and uncertainties in the different samples. The number of Cepheids, 
the mean and the $\sigma$ of the Gaussian fit are also labelled. \label{fig:cepdistr}}
\end{figure*}

\begin{figure*}
\centering
\includegraphics[width=9cm]{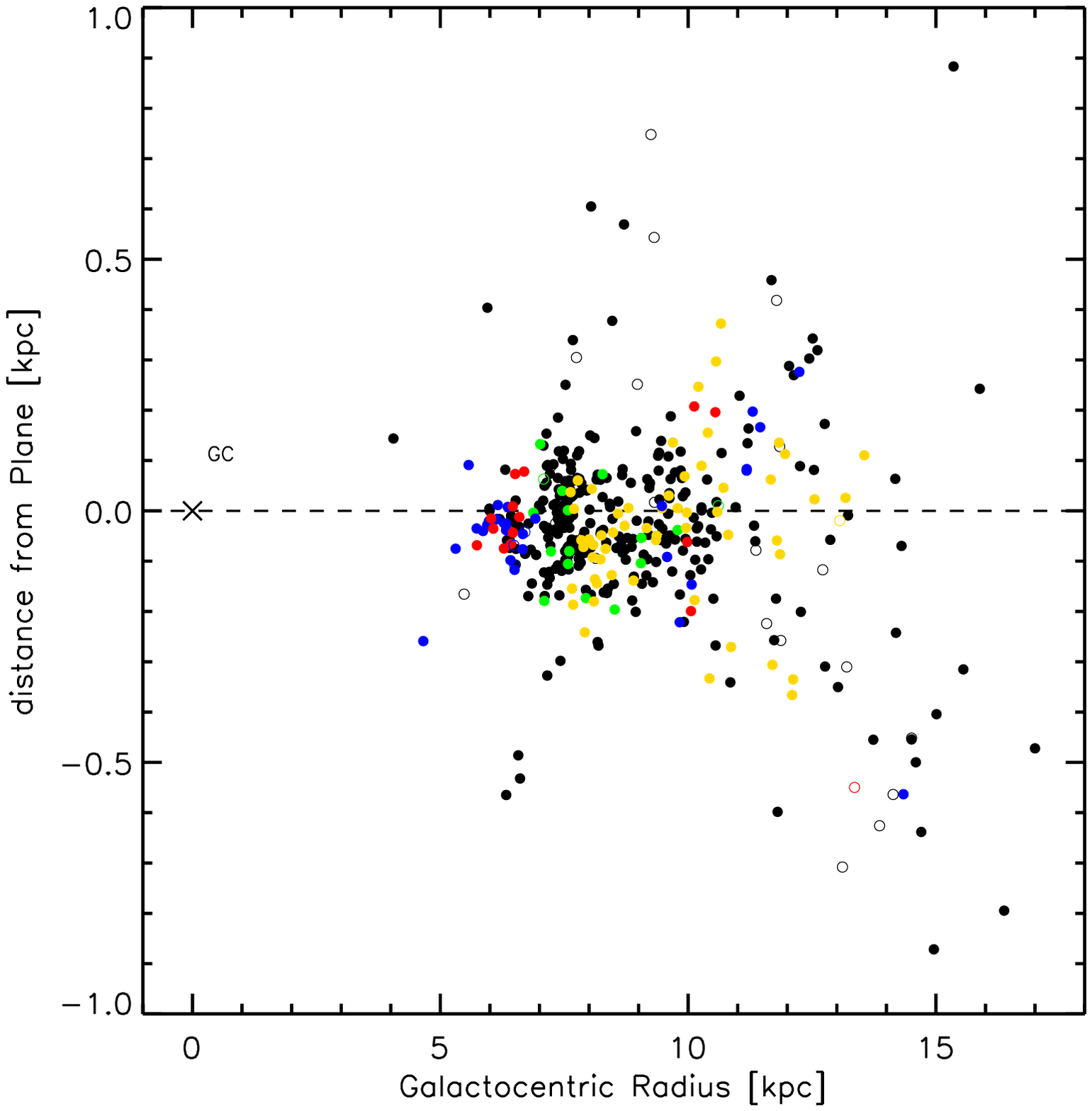}
\caption{
Distance from the Galactic plane of Galactic Cepheids with accurate iron 
abundances versus Galactocentric distance. Symbols and colors are the same as 
in Fig.~2, the black cross marks the Galactic Center (GC).}
\label{fig:cepazim}
\end{figure*}

\clearpage

\begin{table*}%
\caption{
Sample of Galactic Cepheids for which the iron abundance was 
measured using high-resolution, high signal-to-noise      
UVES spectra.}
\label{tab:tab_distances}
\centering
\begin{tabular}{r r ccc c cc ll}
\hline
\hline
Name  &
log($P$)  & $<J>$  &
$<H>$  &
$<K>$  &
$[Fe/H]$  &
$\mu$$^e$  &
$\Rg$$^f$  &
\multicolumn{2}{c}{Notes} \\
  &
[days] &
mag  &
mag  &
mag  &
  &
mag &
(pc) &  \\
\hline
  V340 Ara        &       1.3183        & 7.302        & 6.754                            & 6.534                                      &  0.53                                 $\pm$   0.09        & 12.99 $\pm$ 0.05 &  4657                        $\pm$    427                        & b $^{\dagger}$  &      \\
    AS Aur        &       0.5017        &  9.90        &  9.46                            &  9.33                                      &  0.00                                  $\pm$  0.08        & 13.18 $\pm$ 0.06 & 12244                        $\pm$   469                                     & c   &     \\
    KN Cen        &       1.5321        & 6.439        & 5.769                            & 5.474                                      &  0.55                                  $\pm$  0.12        & 12.51 $\pm$ 0.05 & 6498                        $\pm$    417                                     & b    &    \\
    MZ Cen        &       1.0151        &  8.44        &  7.87                            &  7.61                                      &  0.27                                  $\pm$  0.10        & 13.01 $\pm$ 0.07 & 6501                        $\pm$   391 & c     &   \\
   OO Cen        &       1.1099        &  8.39        &  7.72                            &  7.45                                      &  0.20                                  $\pm$  0.06        & 13.08 $\pm$ 0.07 & 6025                        $\pm$   389 & c      & \#$^d$    \\
    TX Cen       &       1.2328        &  7.17        &  6.66                            &  6.40                                      &  0.44                                  $\pm$  0.13        & 12.58 $\pm$ 0.07 & 6070                        $\pm$    419 & c      & \#    \\
  V339 Cen        &       0.9762        & 6.354        & 5.877                            & 5.690                                      &  0.06                                  $\pm$  0.03        & 11.07 $\pm$ 0.05 & 6917                        $\pm$    446                                     & b     &   \\
    VW Cen        &       1.1771        & 7.588        & 7.028                            & 6.805                                      &  0.41                                  $\pm$  0.08        & 12.78 $\pm$ 0.05 & 6417                        $\pm$    405                                     & b    &    \\
    RW CMa       &       0.7581        &  8.43        &  7.96                            &  7.75                                      & -0.07                                  $\pm$  0.08        & 12.38 $\pm$ 0.07 & 10057                        $\pm$   445                                     & c  &   \#      \\
    SS CMa        &       1.0921        & 7.370        & 6.863                            & 6.664                                      &  0.06                                  $\pm$  0.04        & 12.40 $\pm$ 0.05 &  9829                        $\pm$   439                                     & b &       \\
    TV CMa        &       0.6693        & 8.022        & 7.582                            & 7.364                                      &  0.01                                  $\pm$  0.07        & 11.72 $\pm$ 0.05 &  9575                        $\pm$   447                                     & a  &      \\
    AA Gem        &       1.0532        & 7.636        & 7.201                            & 7.048                                      & -0.08                                  $\pm$  0.05        & 12.74 $\pm$ 0.05 & 11454                        $\pm$   459                                     & a  &      \\
    BW Gem        &       0.4208        &  9.83        &  9.30                            &  9.16                                      & -0.22                                  $\pm$  0.09        & 12.65 $\pm$ 0.07 & 11302                        $\pm$   463                                     & c  &      \\
    FT Mon        &       0.6843        & 10.400        & 9.937                            & 9.754                                      & -0.13                                 $\pm$   0.08        & 14.16 $\pm$ 0.05 & 14344                        $\pm$   468                        & b $^{\dagger}$  &      \\
    SV Mon        &       1.1828        & 6.249        & 5.829                            & 5.666                                      &  0.12                                  $\pm$  0.08        & 11.80 $\pm$ 0.05 & 10070                        $\pm$   453                                     & a  &      \\
    TY Mon        &       0.6045        & 9.302        & 8.865                            & 8.668                                      &  0.02                                  $\pm$  0.08        & 12.82 $\pm$ 0.05 & 11180                        $\pm$   451                                     & a  &      \\
    TZ Mon        &       0.8709        & 8.447        & 8.005                            & 7.795                                      & -0.02                                 $\pm$   0.07        & 12.83 $\pm$ 0.05 & 11183                        $\pm$   451                        & a   &      \\
    GU Nor        &       0.5382        &  7.61        &  7.17                            &  6.97                                      &  0.08                                  $\pm$  0.06        & 10.90 $\pm$ 0.07 & 6663                        $\pm$    450                                     & c  &      \\
    IQ Nor       &       0.9159        &  6.79        &  6.24                            &  6.02                                      &  0.22                                  $\pm$  0.07        & 11.12 $\pm$ 0.08 & 6691                        $\pm$    448                                     & c  &   \#      \\
    QZ Nor       &       0.5782        & 7.085        & 6.748                            & 6.614                                      &  0.19                                  $\pm$  0.08        & 11.53 $\pm$ 0.05 & 6283                        $\pm$    447                                     & b  &  \#      \\
    RS Nor       &       0.7923        &  7.32        &  6.85                            &  6.64                                      &  0.18                                  $\pm$  0.08        & 11.39 $\pm$ 0.07 & 6385                        $\pm$    449                                     & c  &   \#      \\
    SY Nor        &       1.1019        & 6.638        & 6.091                            & 5.864                                      &  0.27                                 $\pm$   0.10        & 11.59 $\pm$ 0.05 & 6286                        $\pm$    446                       & b $^{\dagger}$    &    \\
    TW Nor        &       1.0329        & 7.442        & 6.712                            & 6.375                                      &  0.27                                  $\pm$  0.10        & 11.67 $\pm$ 0.05 & 6160                        $\pm$    447                                     & b   &     \\
  V340 Nor        &       1.0526        & 6.211        & 5.745                            & 5.573                                      &  0.07                                  $\pm$  0.07        & 11.22 $\pm$ 0.05 & 6483                        $\pm$    449                                     & b    &    \\
    RS Ori        &       0.8789        & 6.398        & 6.020                            & 5.860                                      &  0.11                                  $\pm$  0.09        & 11.00 $\pm$ 0.05 &  9470                        $\pm$   453                                     & a   &     \\
    BM Pup       &       0.8572        &  8.42        &  7.93                            &  7.76                                      & -0.07                                  $\pm$  0.08        & 12.74 $\pm$ 0.07 &  9981                        $\pm$   435                                     & c  & \#       \\
    CK Pup       &       0.8703        & 10.273        & 9.703                            & 9.456                                      & -0.12                                 $\pm$   0.08        & 14.37 $\pm$ 0.05 & 13357                        $\pm$   423                        & b $^{\dagger}$ & \#      \\
    WY Pup       &       0.7202        & 8.945        & 8.577                            & 8.453                                      & -0.10                                 $\pm$   0.08        & 13.09 $\pm$ 0.05 & 10549                        $\pm$   430                        & b $^{\dagger}$ & \#       \\
    WZ Pup       &       0.7013        &  8.71        &  8.33                            &  8.20                                      & -0.07                                  $\pm$  0.06        & 12.76 $\pm$ 0.07 & 10123                        $\pm$   437                                     & c  &  \#      \\
    KQ Sco        &       1.4577        & 5.945        & 5.229                            & 4.924                                      &  0.52                                  $\pm$  0.08        & 11.67 $\pm$ 0.05 & 5948                        $\pm$    451                                     & b  &      \\
    RY Sco        &       1.3078        & 4.930        & 4.379                            & 4.134                                      &  0.06                                  $\pm$  0.02        & 10.54 $\pm$ 0.05 & 6663                        $\pm$    453                                     & b   &     \\
  V470 Sco       &       1.2112        &  6.01        &  5.31                            &  4.98                                      &  0.16                                  $\pm$  0.06        & 10.89 $\pm$ 0.07 & 6461                        $\pm$    454                                     & c  &  \#      \\
    EV Sct        &       0.4901        & 7.608        & 7.184                            & 7.018                                      &  0.09                                  $\pm$  0.07        & 11.54 $\pm$ 0.05 & 6135                        $\pm$    449                                     & b  &      \\
    RU Sct        &       1.2945        & 5.891        & 5.287                            & 5.034                                      &  0.16                                  $\pm$  0.05        & 11.35 $\pm$ 0.05 & 6361                        $\pm$    449                                     & a   &     \\
    UZ Sct        &       1.1686        & 7.405        & 6.749                            & 6.461                                      &  0.45                                 $\pm$   0.07        & 12.29 $\pm$ 0.05 & 5309                        $\pm$    448                        & a   &     \\
  V367 Sct        &       0.7989        & 7.605        & 6.955                            & 6.651                                      & -0.04                                  $\pm$  0.04        & 11.24 $\pm$ 0.05 & 6332                        $\pm$    451                                     & b  &      \\
     X Sct       &       0.6230        & 7.378        & 6.952                            & 6.768                                      &  0.12                                  $\pm$  0.09        & 11.00 $\pm$ 0.05 & 6464                        $\pm$    452                                     & a  &   \#    \\
     Z Sct        &       1.1106        & 6.949        & 6.477                            & 6.263                                      &  0.18                                  $\pm$  0.09        & 12.08 $\pm$ 0.05 & 5733                        $\pm$    445                                     & a   &     \\
    AA Ser        &       1.2340        & 7.563        & 6.773                            & 6.457                                      &  0.38                                  $\pm$  0.21        & 12.39 $\pm$ 0.05 & 5572                        $\pm$    437                                     & a  &      \\
    CR Ser       &       0.7244        & 7.340        & 6.759                            & 6.485                                      &  0.12                                  $\pm$  0.08        & 10.89 $\pm$ 0.05 & 6510                        $\pm$    452                                     & a &   \#       \\
    AV Sgr        &       1.1879        & 6.878        & 6.089                            & 5.730                                      &  0.35                                 $\pm$   0.17        & 11.49 $\pm$ 0.05 & 5980                        $\pm$    455                        & b $^{\dagger}$ &       \\
    AY Sgr       &       0.8175        & 7.127        & 6.531                            & 6.264                                      &  0.11                                  $\pm$  0.06        & 10.97 $\pm$ 0.05 & 6429                        $\pm$    452                                     & a  &  \#      \\
 V1954 Sgr       &       0.7909        &  7.81        &  7.33                            &  7.10                                      &  0.24                                  $\pm$  0.10        & 11.82 $\pm$ 0.07 & 5687                        $\pm$    456                                     & c  &  \#     \\
   V773 Sgr       &       0.7596        &  7.52        &  6.70                            &  6.35                                      &  0.11                                  $\pm$  0.06        & 10.65 $\pm$ 0.07 & 6595                        $\pm$    454                                     & c &   \#       \\
    VY Sgr        &       1.1322        & 7.156        & 6.385                            & 6.042                                      &  0.43                                 $\pm$   0.14        & 11.63 $\pm$ 0.05 & 5862                        $\pm$    453                        & b $^{\dagger}$   &    \\
    WZ Sgr        &       1.3394        & 5.255        & 4.752                            & 4.536                                      &  0.35                                  $\pm$  0.08        & 11.10 $\pm$ 0.05 & 6326                        $\pm$    453                                     & a   &     \\
    XX Sgr        &       0.8078        &  6.44        &  5.97                            &  5.75                                      & -0.07                                  $\pm$  0.07        & 10.55 $\pm$ 0.07 & 6706                        $\pm$    453 & c   &     \\

    \vspace{5 mm}
\end{tabular}
\tablefoot{From left to right: target name, period, \textit{J}, \textit{H}, \textit{K} mean magnitudes, 
iron abundance, distance modulus and Galactocentric distance. 
Notes on the photometric data: 
\textbf{(a)} mean magnitudes provided by \cite{Monson2011} transformed into the 
2MASS photometric system using the relations given in their Table~1.  
\textbf{(b)} Mean magnitudes provided by \cite{Laney1992} and Laney private comm. transformed into the 2MASS 
photometric system using the relations given by \cite{Koen2007}. The objects 
marked with a $^{\dagger}$ (Laney private comm.) do not have a complete coverage of the light-curve (the number of phase points ranges from 4 to 14). 
\textbf{(c)} Mean magnitudes based on single-epoch measurements from the 
2MASS catalogue and the NIR template light curves provided by 
\cite{Soszynski2005}. 
\textbf{(d)} Classical Cepheids for which the spectroscopic iron abundance 
is measured for the first time.
\textbf{(e)} The weighted mean true distance moduli. The errors account for uncertainties 
affecting mean magnitudes and for the intrinsic dispersion of the adopted NIR PW relations.
\textbf{(f)} The weighted mean Galactocentric distances were estimated assuming 
$\Rg$=7.94$\pm$0.37$\pm$0.26 kpc \citep{Groenewegen2008}. The errors account for uncertainties 
affecting both the solar Galactocentric distance and the heliocentric distances.}
\end{table*}


\end{document}